\documentclass[aps,prd,twocolumn,showpacs,nofootinbib,preprintnumbers]{revtex4}

\usepackage{color,amsmath,amssymb,subfigure}
\usepackage[dvips]{graphicx}
\usepackage{ulem}
\usepackage{url}

\newcommand{\sigmav}{\ensuremath{\langle\sigma v\rangle}}
\newcommand{\dnde}{\ensuremath{\frac{dN_\gamma}{dE_{\gamma}}}}

\begin{document}

\title{A generic method to constrain the dark matter model parameters 
from Fermi observations of dwarf spheroids}

\author{
Yue-Lin Sming Tsai$^1$, Qiang Yuan$^{2,3}$, Xiaoyuan Huang$^4$
}

\affiliation{
$^1$National Center for Nuclear Research, Hoza 69, 00-681 Warsaw, Poland\\
$^2$Key Laboratory of Particle Astrophysics, Institute of High Energy 
Physics, Chinese Academy of Sciences, Beijing 100049, P. R. China\\
$^3$Key Laboratory of Dark Matter and Space Astronomy, Purple Mountain 
Observatory, Chinese Academy of Sciences, Nanjing 210008, P. R. China\\
$^4$National Astronomical Observatories, Chinese Academy of Sciences, 
Beijing 100012, P. R. China
}

\date{\today}

\begin{abstract}

Observation of $\gamma$-rays from dwarf galaxies is an effective way
to search for particle dark matter. Using 4-year data of Fermi-LAT
observations on a series of Milky Way satellites, we develop a
general way to search for the signals from dark matter annihilation
in such objects. Instead of giving prior information about the energy 
spectrum of dark matter annihilation, we bin the Fermi-LAT data into
several energy bins and build a likelihood map in the ``energy bin
- flux'' plane. The final likelihood of any spectrum can be easily 
derived through combining the likelihood of all the energy bins.
It gives consistent result with that directly calculated using
the Fermi Scientific Tool. This method is very efficient for the 
study of any specific dark matter models with $\gamma$-rays.
We use the new likelihood map with Fermi-LAT 4 year data to fit
the parameter space in three representative dark matter models: 
i) toy dark matter model, ii) effective dark matter operators, 
and iii) supersymmetric neutralino dark matter.

\end{abstract}

\maketitle

\section{Introduction}

Detecting particle dark matter (DM) is one of the most important
task for the modern physics, although its existence has been revealed
through astronomical observations for more than seventy years.
The DM particles could annihilate themselves to produce standard
model (SM) particles such as electrons/positrons, proton/antiprotons,
$\gamma$-ray photons and neutrinos, which provides us a chance to
detect the DM indirectly in the cosmic radiation. The Fermi Large 
Area Telescope (Fermi-LAT) is currently the best detector for 
$\gamma$-ray detection in space and could substantially increase 
the sensitivity for DM search.

The dwarf spheroidal satellites (dSphs) in the Milky Way are generally 
regarded as ideal candidates when searching for possible DM annihilation 
signals due to their high DM content and low baryonic contamination. 
The search for $\gamma$-ray emission from dSphs with Fermi-LAT data
was performed in many works \cite{Abdo:2010ex,Ackermann:2011wa,
GeringerSameth:2011iw,Cholis:2012am,GeringerSameth:2012sr,
Mazziotta:2012ux,Baushev:2012ke,Huang:2012yf}. 
Without finding any significant ``signal'' from these targets, stringent
upper limits on DM annihilation cross section can be derived. It
was shown that for $m_\chi\lesssim20$ GeV the canonical thermally
produced DM with cross section $\sim 3\times10^{-26}$ cm$^3$ s$^{-1}$
was ruled out with two-year Fermi-LAT observations on 10 dSphs
\cite{Ackermann:2011wa} (see also \cite{GeringerSameth:2011iw} for
similar conclusion). 

There are usually two ways to extract the limits on DM models from
the $\gamma$-ray observations. The most conservative and 
DM-model-independent way is to require the DM-induced signal not to 
exceed the flux/limit of specific source. But in this way, the spectral 
and spatial information of DM is missing and the constraint is usually 
weak\footnote{Note in \cite{Mazziotta:2012ux} a stronger constraint was
given. In that work a lower upper limit of $\gamma$-ray flux was derived
through subtracting the photon counts of choosen ``background'' region
from the ``signal'' region.}. Another way which could be more realistic 
is to incorporate the expected $\gamma$-ray spectrum and/or the sky map 
in the analysis. However, it needs to input the model of DM when analyzing 
the data and is not convenient for generic purpose. 

In this work we alternatively adopt 
a more general way to extract the limits on the $\gamma$-ray
emission of the dSphs from Fermi-LAT data. The Fermi-LAT observational
data are binned into a series of energy bins, and the likelihood in 
each energy bin is calculated assuming a constant value of the flux 
from DM annihilation in this energy bin. Given any shape of the 
$\gamma$-ray spectrum, we can easily derive the total likelihood 
with such a likelihood map on ``energy-flux'' plane. This method is
model-independent in the level of extracting the likelihood of 
$\gamma$-ray flux in each energy bin, and it is more flexible
and time-saving to subsequently discuss any kind of particle models. 
The detailed description of the method is given in Sec. II. In Sec. III we apply 
this method to several DM models.

\section{Likelihood Map}

In this section, we will describe the methodology of how to build a 
particle model-independent likelihood map. The $\gamma$-ray flux
from annihilation of DM in a dSph is
\begin{equation}
\phi(E)=\frac{\sigmav}{8\pi m_{\chi}^2}\times \dnde \times J,
\label{eq:gammaflux}
\end{equation}
where $m_\chi$ is the mass of DM particle, $\sigmav$ is the average
velocity-weighted annihilation cross section, $\dnde$ is the $\gamma$-ray
yield spectrum of one annihilation of a DM pair, and $J=\int
dl\,d\Omega\,\rho^2(l)$ is the integral density square of DM.

From Fermi-LAT 2-year result \cite{Ackermann:2011wa}, we can see 
$\sigmav$ with the $b\bar{b}$ channel for $m_\chi\lesssim 25$ GeV 
is lower than the result expected from thermal equilibrium, 
$\sigmav\sim 3\times 10^{-26}\,\rm{cm}^3 s^{-1}$. It could be a hint 
that Fermi-LAT can rule out DM annihilation to $b\bar{b}$, 
at least, at the lower mass region, $m_{\chi}\lesssim25$ GeV.   
However, the above information cannot be true for realistic DM 
models. In most of particle models such as supersymmetry (SUSY), 
$\gamma$-rays can be produced via several annihilation channels. 
Moreover, $\dnde$ can differ from one point to another in the SUSY 
parameter space. Therefore it will be more useful to have a likelihood 
map which does not depend on a pre-determined $\dnde$. 
To do so, we bin the photons into small energy bins $[E_i,E_{i+1}]$, 
and assume the differential energy spectrum $\dnde$ in the small energy 
interval to be constant $C_i$. Then the likelihood is built based on two 
new variables, $E_i$ and $\phi_{ij}=\frac{\sigmav}{8\pi m_{\chi}^2}
\times C_i\times J_j$, where $i$ is the index of energy bin and $j$ is 
the index of individual dSph. We adopt the same 10 dSphs in this study 
as in \cite{Ackermann:2011wa}. 

To calculate the likelihood from Fermi-LAT data we employ $4$-year 
Fermi-LAT data\footnote{http://fermi.gsfc.nasa.gov/ssc/data} recorded 
from 4 August 2008 to 2 August 2012, with the pass 7 photon selection.
The LAT Scientific Tools version v9r27p1 are used for the analysis.
The ``SOURCE'' (evclass=2) event class is selected, and the
recommended filter cut ``(DATA\_QUAL==1) \&\& (LAT\_CONFIG==1)
\&\& ABS(ROCK\_ANGLE)$<52$'' is applied. The photon energy range 
is choosen from $200$ MeV to $200$ GeV, and the region-of-interest 
(ROI) is adopted to be a $14^{\circ}\times14^{\circ}$ box centered 
around the center of each dSph. The instrument response function used 
is ``{\tt P7SOURCE\_V6}''. For the diffuse background, we use the 
Galactic diffuse model {\tt gal\_2yearp7v6\_v0.fits} and the isotropic
diffuse spectrum {\tt iso\_p7v6source.txt} provided by the Fermi
Science Support
Center\footnote{http://fermi.gsfc.nasa.gov/ssc/data/access/lat/Background-
Models.html}. The point sources of the second LAT source catalog are
also included in the modeling \cite{Fermi:2011bm}. The binned likelihood
method is used to calculate the likelihood $L^{\rm LAT}_{ij}$. 
The DM contribution is modelled as a point
source located at the central position of each dSph. In the analysis the 
free parameters include all the nomalizations of the second LAT sources 
in the ROI, and the normalizations of the two diffuse backgrounds.

We follow the method described in \cite{Ackermann:2011wa} to combine the 
results of different dSphs. Note that $J_j$ factor is absorbed in 
$\phi_{ij}$ when calculating $L^{\rm LAT}_{ij}$. To get rid of the 
effect of $J_j$, we define 
\begin{equation}
\label{eq:phij}
\psi_{i}=\phi_{ij}/J_j=\frac{\sigmav}{8\pi m_{\chi}^2}\times C_i,
\end{equation}
and write the joint likelihood function as
\begin{align}\label{eq:L}
L_i(D| \psi_i&)=\prod_{j} L^{\rm LAT}_{ij}
(D|\psi_{i}, \mathbf{p}_j)\nonumber\\
 & \times \frac{1}{\ln(10)\,J_j \sqrt{2\pi}\sigma_{j}} 
e^{-\left[\log_{10}(J_j)-\overline{\log_{10}
({J}_j)}\right]^2/2\sigma_{j}^2}\; ,
\end{align}
where $D$ represents the binned $\gamma$-ray data and $\{\mathbf{p}\}_j$ 
are the ROI-dependent model parameters. $\overline{\log_{10}(J_j)}$ and 
$\sigma_j$ are the mean and standard deviations of the Gaussian 
distribution of $\log_{10}{(J_j)}$. 
It is worth pointing out that the $J$-factors are weakly dependent on 
the density profile of the halo. Taking Draco as an example, the $J$
factor for Burkert profile \cite{Cholis:2012am} according to the 
stellar kinematics \cite{Salucci:2011ee} is similar with that
for NFW profile as given in \cite{Ackermann:2011wa}. Moreover, the authors 
of Ref. \cite{Charbonnier:2011ft} perform a scan with five halo parameters 
varied. Interestingly, they found that the highly cuspy NFW profile 
and a flat isothermal core profile could fit the data equally well 
and the posterior distribution of $J$ was still similar with the value 
given in \cite{Ackermann:2011wa}. In this work we take the values of 
$J$-factors and the uncertainties from Table~I of \cite{Ackermann:2011wa}.

\begin{figure}[t!]
\centering
\includegraphics[width=1.1\columnwidth]{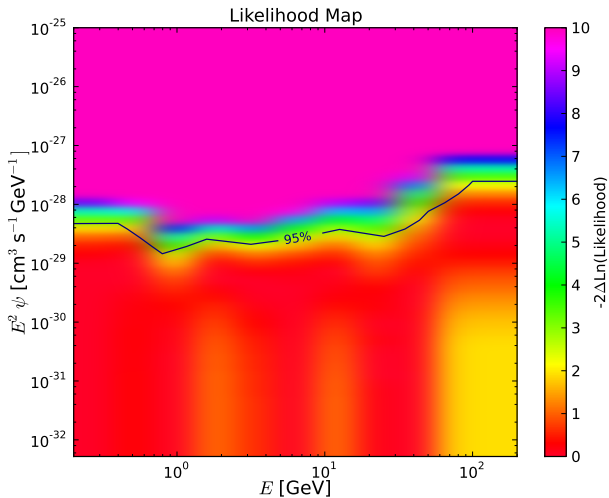}
\caption{\label{fig:like_map} 
The likelihood map on ($E_\gamma$, $E_\gamma^2\psi$) plane based on
4-year Fermi-LAT data on the 10 dSphs. The black solid line is one-sided 
95\% exclusion limit. The color shows the value of $-2\Delta 
\ln\mathcal{L}$ with the likelihood normalized in each energy bin.
}
\end{figure}

We preform a grid scan with $10$ energy bins logarithmically spaced between 
$0.2$ and $200$ GeV, and $50$ bins of $\psi_i$ logarithmically spaced from 
$10^{-33}$ to $10^{-25}$ cm$^3$ s$^{-1}$ GeV$^{-3}$. For each given 
($E_i$, $\psi_i$) point, we vary all the nuisance parameters as well 
as $J_i$, known as the ``profile likelihood'' technique \cite{Rolke:2004mj}, 
and take the maximum likelihood of Eq. (\ref{eq:L}) as the final likelihood 
probability value. The combined likelihood map based on $E_i$ and 
$E_i^2\psi_i$ is shown in Fig. \ref{fig:like_map}. For each energy bin, 
we normalize the maximum likelihood to one. The solid line is a one-sided 
95\% confidence level, corresponding to $-2\Delta \ln\mathcal{L} =2.71$. 
One should bear in mind that these limits shown in Fig. 
\ref{fig:like_map} are totally independent of the theoretical model 
parameters of DM, such as $m_{\chi}$, $\sigmav$, and $\dnde$. 
It indicates that any expected $\gamma$-ray flux should not exceed the 
limits in the whole energy range from 0.2 GeV to 200 GeV. However, such 
a requirement is too conservative compared with the combined 
likelihood of all energy bins (see the next paragraph). If one wants 
to discuss the specific models of the $\gamma$-ray emission and derive 
the constraints on the specific model parameters such as 
$(m_{\chi},\sigmav)$ of DM annihilation, the spectral information will 
be required. However, such a likelihood map in ($m_{\chi}$, $\sigmav$) 
plane is $\dnde$ dependent.

\begin{figure}[t!]
\centering
\includegraphics[width=1.0\columnwidth]{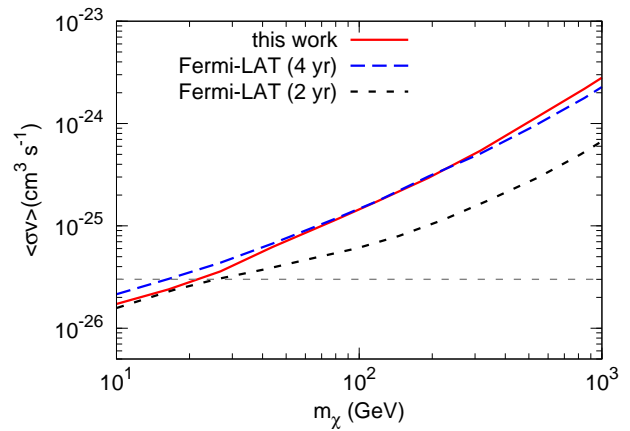}
\caption{\label{fig:sv_comp} 
Comparison of the results using the likelihood map method developed in
this work (solid) and that derived using Fermi Scientific Tool 
(long dashed). Here we assume $b\bar{b}$ channel of DM annihilation 
and the results from the ten dSphs are combined. The 2-year result from 
\cite{Ackermann:2011wa} is also shown. 
}
\end{figure}

Combining the likelihood of each energy bin we can get the total likelihood 
for any input spectrum \dnde. The total likelihood can be calculated as
\begin{equation}\label{eq:Ltot}
L=\prod_{i} L_i(D|\psi_i),
\end{equation}
where $\psi_i$ is the expected flux of the spectrum in the $i$th energy 
bin, with $C_i\approx\left.\dnde\right|_{\sqrt{E_iE_{i+1}}}$.
Inserting Eq. (\ref{eq:phij}) to (\ref{eq:Ltot}), one can translate 
the likelihood map from $(E_i,\,\psi_i)$ plane (Fig. \ref{fig:like_map}) 
to ($m_{\chi}$, $\sigmav$) plane. As mentioned above, the likelihood
map in ($m_{\chi}$, $\sigmav$) plane will depend on the spectrum $\dnde$.

At the level of detailed DM model, $m_{\chi}$, $\sigmav$ and $\dnde$ 
are variable with other intrinsic parameters of the model. With the above
likelihood map (Fig. 1), we can easily find the likelihood of any set
of DM parameters. Of course, for each model point one can implement the
output $\gamma$-ray spectrum in the \texttt{Fermi Scientific tools}
and calculate its likelihood. But it will be much more computing 
time-consuming compared with our method.

In Fig. \ref{fig:sv_comp} we give the comparison of our result with
that derived using the Fermi tool (following directly the way in 
\cite{Ackermann:2011wa} but using 4-yr data), for DM annihilation
to $b\bar{b}$ channel. We can see that these two results agree well 
with each other. Also the two-year result given in \cite{Ackermann:2011wa} 
is shown. The results show that the 4-yr constraint is even weaker than 
that of 2-yr constraint. Similar result was also reported recently 
\cite{Drlica-Wagner2012}, and it was possibly due to the update to pass 
7 of the data and the statistical fluctuations.
One should not be confused with the 95\% upper limit in Fig. 
\ref{fig:sv_comp} and Fig. \ref{fig:like_map}. The line in Fig. 
\ref{fig:like_map} is 95\% upper limit of $E_i^2\psi_i$ at given 
$\gamma$-ray energy bin, while the line in Fig. \ref{fig:sv_comp} is 
95\% upper limit of $\sigmav$ at given $m_{\chi}$ based on total 
likelihood for $\gamma$-rays from 0.2 to 200 GeV. A model point 
disfavoured by likelihood at some certain energy bin must be also 
ruled out by the total likelihood. On the contrast, a model point 
disfavoured by the total likelihood function does not mean that it is
necessarily ruled out by the 95\% line of Fig. \ref{fig:like_map}.

\section{Particle model fit}

In this section we apply the method described in Sec. II to specific
DM models to derive the constraints. Three particle models, a toy model, 
the effective DM model, and the supersymmetry neutralino DM are
adopted as examples.

\subsection{Toy model}

To generalize the upper limit on $m_\chi-\sigmav$ plane, we design a toy 
DM model which annihilates to $\gamma$-rays only via $b\bar{b}$, 
$\tau^+\tau^-$, and $W^+W^-$ channels. We only consider these three 
channels because the differences between them are relatively large
compared with other channels. 
 
With the 4-year likelihood map (Fig. \ref{fig:like_map}), we conduct a 
random scan of the four parameters, $m_{\chi}$, $\sigmav_{b\bar{b}}$, 
$\sigmav_{\tau^+\tau^-}$, and $\sigmav_{W^+W^-}$. The total cross section 
$\sigmav$ equals to the sum of the three channels. The best-fit point 
obtained in the scan is: $m_{\chi}=200$ GeV, $\sigmav=2.39\times 10^{-25} 
{\rm cm^3\,s^{-1}}$, ${\rm BR}_{b\bar{b}}=0.47$, ${\rm BR}_{\tau^+\tau^-}
=0.53$, and ${\rm BR}_{W^+W^-}=10^{-3}$. Not surprisingly, the mixed 
channels can give better fit due to more degrees of freedom. 
However, the current Fermi-LAT data can not effectively
constrain the branching ratios, namely the shapes of $\dnde$.


\begin{figure}[t!]
\centering
\includegraphics[width=1.0\columnwidth]{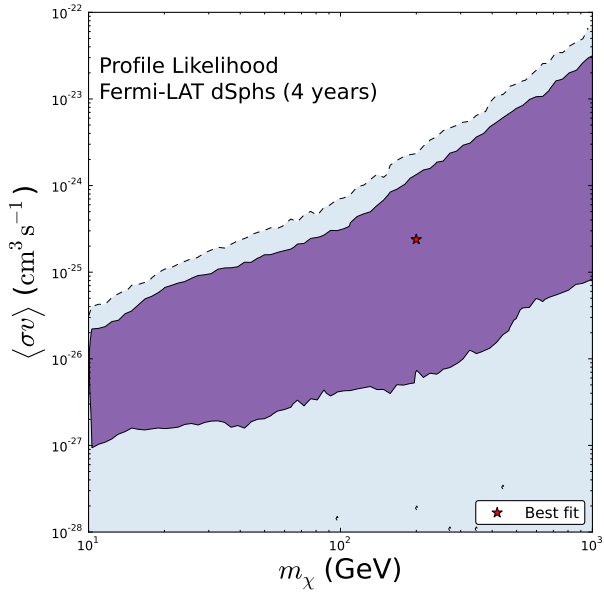}
\caption{\label{fig:data_driven} 
The 68\% (inner) and 95\% (outer) profile likelihood contour for the 
toy model.
}
\end{figure}

In Fig. \ref{fig:data_driven}, we present 68\% (inner) and 95\% (outer) 
contours on the $m_{\chi}-\sigmav$ using the profile likelihood method. 
Since we project the model parameter space from 4 to 2 variables, 
the 68\% and 95\% contours are defined with $\int_{R} L(D|m_\chi,\sigmav) 
\,{\rm d}m_{\chi}\,{\rm d}\sigmav$ equals to 0.68 and 0.95, respectively.  
Here, $R$ is the region within the contours. Compared with Fig. 
\ref{fig:sv_comp}, we can clearly see that the 95\% upper limit of this 
toy model is significantly higher than that of $b\bar{b}$ channel only
for $m_{\chi}\gtrsim 100$ GeV and slightly higher for lower masses. 
This is because of the inclusion of $\tau^+\tau^-$ channel in the toy
model. According to the results in \cite{Ackermann:2011wa}, 
for $m_{\chi}\lesssim 100$ GeV the constraints on $\sigmav_{b\bar{b}}$
and $\sigmav_{\tau^+\tau^-}$ are comparable, while for higher DM mass
the constraint on $\sigmav_{\tau^+\tau^-}$ becomes much weaker than
that on $\sigmav_{b\bar{b}}$. Including $W^+W^-$ channel also makes the
total constraint weaker.

\subsection{Effective DM models}

\begin{figure}[t!]
\centering
\includegraphics[width=1.0\columnwidth]{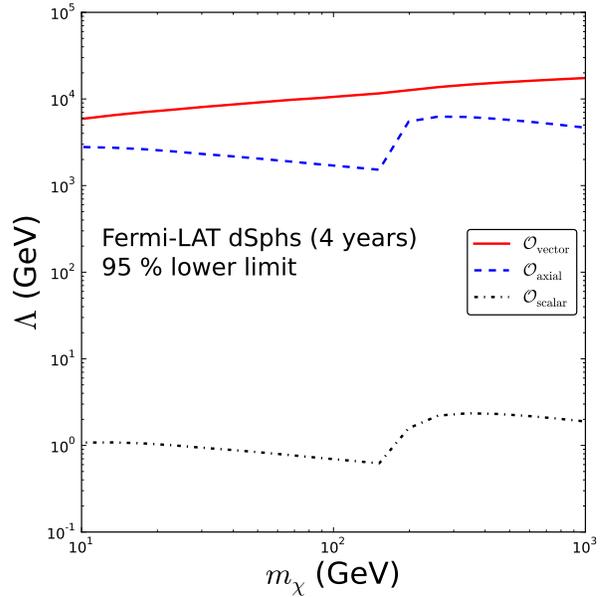}
\caption{\label{fig:effDM} 
The 95\% lower limits on $\Lambda$ for each operator from the 4-year 
Fermi-LAT dSph data.
}
\end{figure}

Before assuming particular DM models, we may adopt an effective interaction 
approach to describe the interactions of the DM particle with the SM 
particles \cite{effective-1,gamma-4,effective-2,gamma-8,Cheung:2012gi,
Mambrini:2012ue}.
In such scenario, the DM particle exists in a hidden sector, which 
communicates with the SM sector via a heavy degree of freedom in the 
connector sector. At energy scale well below this heavy mediator 
we can introduce a new energy scales $\Lambda$ to describe the effective 
couplings between DM and SM fermions. In this study, we consider the 
interactions between fermionic DM field $\chi$ and SM fermion $f$ 
described by scalar ($\mathcal{O}_{\rm{scalar}}$), vector 
($\mathcal{O}_{\rm{vector}}$) or axial ($\mathcal{O}_{\rm{axial}}$) operators 
\begin{eqnarray}
\mathcal{O}_{\rm{vector}} & = & \sum_f \frac{C_V^f}{\Lambda_V^2} \left( \bar \chi \gamma^\mu \chi \right) 
\left( \bar f \gamma_\mu f \right) \; ,\\
\mathcal{O}_{\rm{axial}} & = & \sum_f \frac{C_A^f}{\Lambda_A^2} \left( \bar \chi \gamma^\mu \gamma^5 \chi \right) 
\left( \bar f \gamma_\mu \gamma^5 f \right) \; ,\\
\mathcal{O}_{\rm{scalar}} & = & \sum_f \frac{C_S^f m_f}{\Lambda_S^3} \left( \bar \chi \chi  \right)
\left( \bar f f \right) \; ,
\end{eqnarray}
where $C_{i=V,A,S}$ is an effective coupling constant of order 
$\mathcal{O}$ that can be absorbed into $\Lambda_i$ and $f$ runs over all 
the lepton and quark states. Except $\mathcal{O}_{\rm{scalar}}$ whose 
coupling also depends on $m_f$, $\mathcal{O}_{\rm{vector}}$ and 
$\mathcal{O}_{\rm{axial}}$ both have universal coupling to all the fermions. 

We follow the calculation in \cite{Cheung:2012gi} to obtain $\sigmav$ 
and $\dnde$ for the effective DM models, and constrain each of the 
effective operators using the 4-year Fermi-LAT dSph data. Since we assume 
a universal coupling for all the fermions and the top quark channel only 
opens at $m_\chi>172$ GeV, the upper limits of $\sigmav$ of these three 
operators at low mass region $m_\chi<172$ GeV behave like $b\bar{b}$ channel 
in Fig. \ref{fig:sv_comp}. Even if the $t\bar{t}$ channel opens, due to the 
large contribution to the $\gamma$-ray flux through the $b\bar{b}$ channel,
the result is still not much different from the one $\chi\chi$ directly 
to the $b\bar{b}$. However, it is worthy to mention that the constraints on
$\sigmav$ of the three operators will depend on $m_{f}$ (see Eqs. (A1), 
(A4) and (A7) in \cite{Cheung:2012gi}). 

In Fig. \ref{fig:effDM}, we show the 95\% lower limits on ($m_{\chi}$, 
$\Lambda$) plane for operators $\mathcal{O}_{\rm{scalar}}$ (black 
dash-dotted), $\mathcal{O}_{\rm{vector}}$ (red solid) and 
$\mathcal{O}_{\rm{axial}}$ (blue dashed) and $\mathcal{O}_{\rm{scalar}}$. 
It is clear to see that both $\mathcal{O}_{\rm{axial}}$ and 
$\mathcal{O}_{\rm{scalar}}$ have a kink at the top quark mass because 
the $t\bar{t}$ channel opens. For operator $\mathcal{O}_{\rm{vector}}$ 
there is only a very weak dependence on $m_f$, and the $t\bar{t}$ 
channel is suppressed. 

It is of interest to compare the limit on ($m_{\chi},\Lambda$) plane 
driven by the Fermi-LAT gamma-ray low-latitude result \cite{Cheung:2012gi} 
and our dSphs lower limits. The limits from dSphs for all three operators 
are slightly stronger. However, comparing our result with the one obtained 
from Galactic radio \cite{Mambrini:2012ue}, we found that the constraint 
for $\mathcal{O}_{\rm{vector}}$ from dSphs can be stronger than the 
Galactic radio. 

\subsection{The Minimal Supersymmetric Standard Model}

Undoubtedly, among various particle physics models, SUSY neutralino is 
the most popular DM candidate. However, many recent published results 
such as a Higgs boson candidate discovery at the LHC, 
flavour physics, $\delta(g-2)_\mu$ (the muon's anomalous 
magnetic moment), and the relic density of DM can set strong constraints 
on the SUSY parameter space. In addition, Fermi-LAT $\gamma$ ray data 
can also test the SUSY DM, e.g., the minimal supersymmetric standard
model (MSSM) \cite{Bergstrom:2010gh,Cotta:2011pm} and the Constrained MSSM 
\cite{Scott:2009jn,Roszkowski:2012uf}. In this subsection, we use 4-year 
Fermi data on dSphs to test the MSSM scenario of DM.

Given the large number of parameters in the MSSM, we have to take 
some reasonable and simplified assumptions. We start by assuming that no 
CP violating phases are present, and that all mass matrices and trilinear
couplings are diagonal in flavor space, in order not to violate the
quite stringent constraints on flavor changing neutral currents (FCNCs).
Moreover, we take the first and second generations of scalars to be
degenerate, again motivated by experimental constraints, such as
$K-\bar{K}$ mixing. As the trilinears are proportional to the fermion 
masses, we can safely ignore the contributions from the first and second
generations as they are dwarfed by the third generation. 

Unless there is a lighter gravitino or R-parity is not conserved, the 
lightest neutralino is the only MSSM particle that can make a good DM 
candidate. The lightest neutralino is the lightest mass eigenstate of a 
mixed gauge-eigenstate, Bino, Wino, up-type Higssino, and down-type Higgsino,
\begin{equation}
\chi^0_{1}=Z_{\rm{bino}} \tilde{B} +Z_{\rm{wino}} \tilde{W}+ 
Z_{H_u} \tilde{H_u} + Z_{H_d} \tilde{H_d}.
\end{equation} 
The coefficients $Z_{i}$ ($i={\rm bino,wino},H_u,H_d$) are 
determined by diagonalizing the neutralino mass matrix.
To describe the neutralino compositions, it is convenient to introduce 
a gaugino fraction, $f_g =Z^{2}_{\rm{bino}}+Z^{2}_{\rm{wino}}$. 
When $f_g$ is close to 1, gauginos will dominate the neutralino, on the 
other hand the neutralino will be higgsino-like if $f_g\sim 0$.

\begin{table}[t]
\begin{tabular}{|c|c|}
\hline
\hline
Parameter & Range \\
\hline\hline
bino mass (GeV)&  $m_1=0.5 m_2$ \\
wino mass (GeV)& $10 <m_2 <4\times 10^3$ \\
gluino mass (TeV)& $0.7<m_3 <5$ \\
top/$\tau$-quark trilinear (TeV)&$-7<A_{t},A_{\tau}<7 $\\
b-quark trilinear (TeV)&$A_{b}=0.5 $\\
pseudoscalar mass (TeV)& $0.2<m_A<4$ \\
$\mu$ parameter (TeV)& $10^{-2}<\mu<4$ \\
3rd gen. squark mass (TeV)& $0.3<m_{\tilde{Q}_3}<4$ \\
stau mass (TeV)& $0.1<m_{\tilde{\tau}}<4$ \\
1st/2nd gen. slepton mass (GeV) & $m_{\tilde{L}_{1,2}}=m_1 +50$ \\
1st/2nd gen. squark mass (TeV) & $m_{\tilde{Q}_{1,2}}=2.5$ \\
ratio of Higgs doublet VEVs & $3<\tan\beta <62$ \\
\hline
\hline
\end{tabular}
\caption{The prior ranges of input parameters over which we perform the 
scan of the MSSM. We make two separate scans, one with flat priors  
and the other one with log priors for all mass parameters. 
In both scans, we adopt a flat prior for $\tan \beta$ and trilinear couplings.
}
\label{table:MSSMparams}
\end{table} 

The input parameters and their prior ranges are shown in Table 
\ref{table:MSSMparams}. After the Higgs candidate was discovered 
at the LHC \cite{CMS:2012gu,ATLAS:2012gk}, the Higgs resonance region, 
$2m_{\chi}\sim m_{h}\sim 126$ GeV, can drive a large $\gamma$-ray 
flux because of small $m_{\chi}$ and large $\sigmav$. Besides the Higgs 
resonance region, one can also have a large $\psi_i$ at the Focus point 
region and A-funnel region with small $m_{\chi}$. We let $m_1=0.5 m_2$ 
to obtain a bino-like neutralino. However, we still allow a Higgsino 
like neutralino, particularly at $m_{\chi}\sim1$ TeV, which is strongly 
disfavoured by $\delta(g-2)_\mu$ constraint. We keep the 1st and 2nd 
generations of squarks as heavy as $\sim 2.5$ TeV and the first two 
generations of slepton masses $m_{\tilde{L}_{1,2}}=m_1 +50$ GeV at SUSY 
scale. 

To conduct the investigation, we use the nested sampling algorithm, 
implemented in \texttt{MultiNest}\cite{multinest} which is incorporated 
into \texttt{BayesFITS} package, with 9000 live points, evidence tolerance 
factor 0.5, and sampling efficiency 0.8. To get rid of prior dependence 
and evenly explore the full parameter space, we make two separate runs, 
one with flat priors and the other with log priors for all the mass 
parameters. For the rest parameters we use only the flat priors. The 
observables included in the total likelihood $\mathcal{L}_{\rm{total}}$
used to constrain the parameter space are Higgs mass 
$m_{h}\sim 126$ GeV, WMAP relic density $\Omega_{\chi} h^2$, 
$\ensuremath{B_s\to\mu^+\mu^-}$, $\ensuremath{ b\rightarrow s \gamma }$, 
$\ensuremath{B_u \rightarrow \tau \nu}$, and $\ensuremath{\Delta M_{B_s}}$. 
For SM precision, we also include $m_{W}$ and $\sin\theta_{\rm{eff}}$.  
The theoretical and experimental errors can be found in Ref. 
\cite{Kowalska:2012gs} and references therein. Regarding to the relic 
density we assume that there is no other DM ingredients except neutralino. 
If the relic density is not dominant by neutralino, the DM fluxes 
must be rescaled by a factor $(\Omega_{\chi}/\Omega_{\rm{WMAP}})^2$ because 
of DM densities. We are only interested that $\mathcal{L}_{\rm{dSphs}}$ is 
improved by adding the rescaling factor but $\psi_i$ is still large enough 
to be detectable by Fermi-LAT. To obtain the above solution, it is only 
possible that $\Omega_{\chi}$ is slightly smaller than 
$\Omega_{\rm{WMAP}}$. Therefore, in the level of scan, we will only 
consider a Gaussian likelihood instead of an upper limit for neutralino 
relic density. In addition, we also check the exclusion bounds obtained 
from the Higgs searches at LEP and the Tevatron, by implementing 
\texttt{HiggsBounds-3.8.0} \cite{Bechtle:2011sb}. Note that we do not 
use $\delta(g-2)_\mu$ constraint because it strongly disfavoured 
for $m_{\chi}\gtrsim 600$ GeV where neutralino is most Higgsino-like.
 
The total $\gamma$-ray spectrum from neutralino annihilation is given by 
\begin{equation}
\dnde=\sum_f\sum_X B_f\frac{dN_f^{\gamma}(X)}{dE}\,,
\end{equation}
where $X$ runs over the contributions of secondary photons, final state 
radiation, virtual internal bremsstrahlung, and monochromatic $\gamma$-ray 
line, $B_f$ is the branching ratio into the annihilation channel $f$. 
The contribution from secondary photons is mainly produced through the 
decay of $\pi^0$ and $K^{0}$. For completeness, we also include the 
monochromatic $\gamma$-ray line contributions because $\chi\chi
\rightarrow\gamma\gamma/\gamma Z$ can slightly contribute to our 
likelihood for $m_{\chi}\lesssim 200$ GeV. The spectrum is calculated
using the \texttt{DarkSUSY-5.0.6} package \cite{darksusy}.

It is possible that the annihilation cross section of DM
in the Milky Way today can be boosted by non-perturbative Sommerfeld 
corrections compared with that in the early Universe, especially for 
$m_{\chi}\gtrsim$ TeV in the MSSM case (e.g., \cite{Hisano:2004ds,
Cirelli:2007xd,Hryczuk:2010zi}). In the MSSM, the lightest neutralino 
with mass $\sim$TeV are mostly Wino- or Higgsino-like, with Sommerfeld 
enhancement factor of order $O(1)$ \cite{Hisano:2004ds,Cirelli:2007xd,
Hryczuk:2010zi}.
On the other hand, $\dnde$ can differ by a factor of several between
different simulation codes \texttt{Pythia6} \cite{Sjostrand:2006za}, 
\texttt{Pythia8} \cite{Sjostrand:2007gs} and \texttt{Herwig} 
\cite{Corcella:2000bw}. To investigate possible source of boost factor 
which could come from either the Sommerfeld enhancement or uncertainties 
of $\dnde$, we can employ a phenomenological boost factor $\mathcal{B}$ 
to boost our SUSY prediction flux $\psi_i^{\rm{SUSY}}$ in order to obtain 
the maximum dSphs likelihood, $\psi_i(\mathcal{L}^{\rm{dSphs}}_{\rm{max}})
=\mathcal{B}\psi_i^{\rm{SUSY}}$.
        
\begin{figure}[t!]
\centering
\includegraphics[width=1.0\columnwidth]{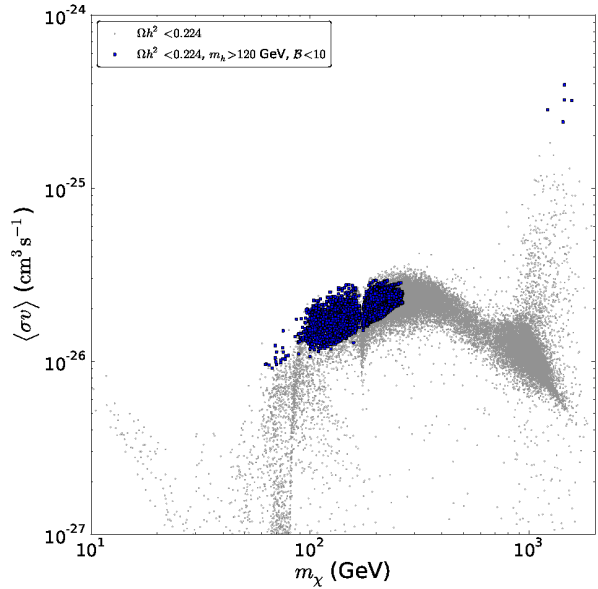}
\caption{\label{fig:mssm_mx_sigv} 
The impact of 4-year Fermi-LAT dSph result on ($m_{\chi}, \sigmav$) plane
of neutralino DM. The grey dots satisfy $\Omega h^2 <0.224$ and 
$-2\ln\mathcal{L}_{\rm{total}}<20$, and the blue squares are selected 
from grey cycle points with $m_{h}>120$ GeV and $\mathcal{B}<10$.
}
\end{figure}

We present the scan result on ($m_{\chi}, \sigmav$) plane in 
Fig. \ref{fig:mssm_mx_sigv}. Totally $\sim 200,000$ sample points are
obtained. The grey dots correspond to the sample surviving from 
$-2\ln\mathcal{L}_{\rm{total}}<20$ and $\Omega h^2 <0.224$ cut. 
From Ref. \cite{Hryczuk:2010zi}, we can see Sommerfeld corrections can 
also reduce the relic density. Since we do not include the Sommerfeld 
corrections in our relic density computation, we here apply a broader 
range $\Omega h^2 <2\Omega_{\rm{WMAP}} h^2$ cut in order to conservatively 
include the possible points. The blue squares are selected from the grey 
dots with criteria $m_{h}>120$ and $\mathcal{B}<10$. We find that the 
blue squares for $m_{\chi}<300$ GeV are mostly bino-like neutralino
but for $m_{\chi}>1200$ GeV they are mostly Higgsino-like neutralino. 
For those Higgsino-like neutralino the boost factor is 
found to be $\mathcal{B}\sim 4$, which could be due to either the Sommerfeld 
effect or the uncertainties of $\dnde$. For the bino-like neutralino with 
lower masses, few points are found with $\mathcal{B}\lesssim 4$. Since the 
Sommerfeld effect does not apply on such small DM mass range and the
uncertainty of $\dnde$ is also only a factor of several, we conclude
that the 4-year Fermi-LAT data is less sensitive to bino-like than 
Higgsino-like neutralino.

\begin{figure}[t!]
\centering
\includegraphics[width=1.1\columnwidth]{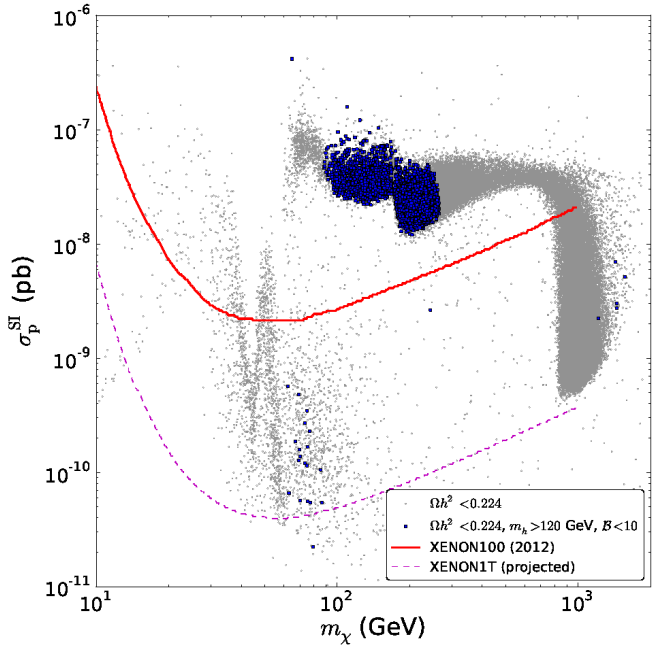}
\caption{\label{fig:mssm_mx_sigsip} 
The scatter plot on the ($m_{\chi}, \sigma_{\rm{p}}^{\rm{SI}}$) plane in 
the MSSM. The color scheme is same as Fig. \ref{fig:mssm_mx_sigv}. 
The solid (red) line is XENON100 (2012) 95\% upper limit and 
the dashed (magenta) line is XENON1T projected sensitivity.   
}
\end{figure}

The constraint on the spin-independent cross section between DM particle
and nucleon $\sigma_{\rm{p}}^{\rm{SI}}$ is shown in Fig. 
\ref{fig:mssm_mx_sigsip}. Also shown are the XENON100 (2012) result
of the direct detection experiment \cite{Aprile:2012nq} and the expected
sensitivity of XENON1T. We notice that XENON100 can exclude most of 
neutralinos which have gaugino fraction, $0.6<f_g<0.9$. However, for 
pure gaugino $f_g\sim 1$ or Higgsino $f_g\sim 0$, Fermi-LAT data still 
can test them prior to XENON1T.

\section{Summary}

In this work we develop a generic method to ensure fast computation of 
the likelihood of any spectral component from Fermi-LAT observations
of dSphs. The Fermi-LAT data of each ROI are binned into several energy 
bins and the likelihood of a point source contribution located at the
ROI center in each energy bin is calculated with the public Fermi 
Scientific Tool. A likelihood map which depends only on the Fermi-LAT
data can be derived without the speculated energy spectrum of the
new component (Fig. \ref{fig:like_map}). Combining the likelihood
of each energy bin we can get the total likelihood of any given energy
spectrum. This method is tested to be in good agreement with the
results directly derived with Fermi Scientific Tool. It is thought
to be of great convenient for the search for DM signals from the
Fermi-LAT data, especially for the discussion of specific DM models.

We apply such a method on several DM models, including a toy model
with mixture of several channels, the effective operator scenarios
and the MSSM models. The Fermi-LAT data can give interesting
constraints on the DM models. First, in our toy model fit, we found 
that the current Fermi-LAT data cannot effectively constrain the 
shapes of $\dnde$. Second, the Fermi-LAT data can improve the lower 
limits of three effective operators on ($m_{\chi},\Lambda$) plane. 
Finally, for the MSSM, we find the Fermi-LAT data are more sensitive 
for the Higgsino-like neutralinos than the bino-like neutralinos. 
Furthermore, Fermi-LAT data can even test the MSSM models with pure 
gaugino or Higgsino compositions more effectively than the current 
direct detection experiments.

The data of the likelihood map and a FORTRAN code \cite{likeCode} to 
calculate the final likelihood of any input spectral function are provided 
for download.

\section*{Acknowledgments}
The authors want to thank Christoph Weniger for
very useful discussions of Fermi-LAT dSphs data. 
YST would like to thank Enrico Sessolo, Leszek Roszkowski 
and Xiao-Jun Bi for the comments on the SUSY model.

This work is supported by National Natural Science Foundation of China
under Grant No. 11105155. QY acknowledges the support from the Key
Laboratory of Dark Matter and Space Astronomy of Chinese Academy of Sciences.
YST was funded in part by the Welcome Programme of the Foundation for 
Polish Science.


\begin{thebibliography}{99}

\bibitem{Abdo:2010ex} 
  A.~A.~Abdo {\it et al.}  [Fermi-LAT Collaboration],
  Astrophys.\ J.\  {\bf 712}, 147 (2010)
  [arXiv:1001.4531 [astro-ph.CO]].

\bibitem{Ackermann:2011wa} 
  M.~Ackermann {\it et al.}  [Fermi-LAT Collaboration],
  Phys.\ Rev.\ Lett.\  {\bf 107}, 241302 (2011)
  [arXiv:1108.3546 [astro-ph.HE]].

\bibitem{GeringerSameth:2011iw} 
  A.~Geringer-Sameth and S.~M.~Koushiappas,
  Phys.\ Rev.\ Lett.\  {\bf 107}, 241303 (2011)
  [arXiv:1108.2914 [astro-ph.CO]].

\bibitem{Cholis:2012am} 
  I.~Cholis and P.~Salucci,
  Phys.\ Rev.\ D {\bf 86}, 023528 (2012)
  [arXiv:1203.2954 [astro-ph.HE]].

\bibitem{GeringerSameth:2012sr} 
  A.~Geringer-Sameth and S.~M.~Koushiappas,
  Phys.\ Rev.\ D {\bf 86}, 021302 (2012)
  [arXiv:1206.0796 [astro-ph.HE]].
  
\bibitem{Mazziotta:2012ux} 
  M.~N.~Mazziotta, F.~Loparco, F.~de Palma and N.~Giglietto,
  arXiv:1203.6731 [astro-ph.IM].
 
\bibitem{Baushev:2012ke} 
  A.~N.~Baushev, S.~Federici and M.~Pohl,
  Phys.\ Rev.\ D {\bf 86}, 063521 (2012)
  [arXiv:1205.3620 [astro-ph.HE]].
  
\bibitem{Huang:2012yf} 
  X.~-Y.~Huang, Q.~Yuan, P.~-F.~Yin, X.~-J.~Bi and X.~-L.~Chen,
  JCAP {\bf 1211}, 048 (2012)
  [arXiv:1208.0267 [astro-ph.HE]].

\bibitem{Fermi:2011bm} 
  [Fermi-LAT Collaboration],
  Astrophys.\ J.\ Suppl.\  {\bf 199}, 31 (2012)
  [arXiv:1108.1435 [astro-ph.HE]].


\bibitem{Salucci:2011ee} 
  P.~Salucci, M.~I.~Wilkinson, M.~G.~Walker, G.~F.~Gilmore, E.~K.~Grebel, A.~Koch, C.~F.~Martins and R.~F.~G.~Wyse,
  Mon.\ Not.\ Roy.\ Astron.\ Soc.\  {\bf 420}, 2034 (2012)
  [arXiv:1111.1165 [astro-ph.CO]].

\bibitem{Charbonnier:2011ft} 
  A.~Charbonnier, C.~Combet, M.~Daniel, S.~Funk, J.~A.~Hinton, D.~Maurin, C.~Power and J.~I.~Read {\it et al.},
  Mon.\ Not.\ Roy.\ Astron.\ Soc.\  {\bf 418}, 1526 (2011)
  [arXiv:1104.0412 [astro-ph.HE]].



\bibitem{Rolke:2004mj} 
  W.~A.~Rolke, A.~M.~Lopez and J.~Conrad,
  Nucl.\ Instrum.\ Meth.\ A {\bf 551}, 493 (2005)
  [physics/0403059].

\bibitem{Drlica-Wagner2012}
A. Drlica-Wagner, Fermi Symposium 2012,


\bibitem{effective-1} 
  J.~-M.~Zheng, Z.~-H.~Yu, J.~-W.~Shao, X.~-J.~Bi, Z.~Li and H.~-H.~Zhang,
  Nucl.\ Phys.\ B {\bf 854}, 350 (2012)
  [arXiv:1012.2022 [hep-ph]].



\bibitem{gamma-4}
J.~Goodman, M.~Ibe, A.~Rajaraman, W.~Shepherd, T.~M.~P.~Tait and H.~-B.~Yu,
  Nucl.\ Phys.\  {\bf B844}, 55-68 (2011)
  [arXiv:1009.0008 [hep-ph]].

\bibitem{gamma-8}
K.~Cheung, P.~-Y.~Tseng and T.~-C.~Yuan,
  JCAP {\bf 1106}, 023 (2011)
  [arXiv:1104.5329 [hep-ph]].

\bibitem{effective-2}
  Z.~-H.~Yu, J.~-M.~Zheng, X.~-J.~Bi, Z.~Li, D.~-X.~Yao and H.~-H.~Zhang,
  arXiv:1112.6052 [hep-ph].


\bibitem{Cheung:2012gi} 
  K.~Cheung, P.~-Y.~Tseng, Y.~-L.~S.~Tsai and T.~-C.~Yuan,
  JCAP {\bf 1205}, 001 (2012)
  [arXiv:1201.3402 [hep-ph]].
  
\bibitem{Mambrini:2012ue} 
  Y.~Mambrini, M.~H.~G.~Tytgat, G.~Zaharijas and B.~Zaldivar,
  JCAP {\bf 1211}, 038 (2012)
  [arXiv:1206.2352 [hep-ph]].




\bibitem{Bergstrom:2010gh} 
  L.~Bergstrom, T.~Bringmann and J.~Edsjo,
  Phys.\ Rev.\ D {\bf 83}, 045024 (2011)
  [arXiv:1011.4514 [hep-ph]].
  
  
\bibitem{Cotta:2011pm} 
  R.~C.~Cotta, A.~Drlica-Wagner, S.~Murgia, E.~D.~Bloom, J.~L.~Hewett and T.~G.~Rizzo,
  JCAP {\bf 1204}, 016 (2012)
  [arXiv:1111.2604 [hep-ph]].





\bibitem{Scott:2009jn} 
  P.~Scott, J.~Conrad, J.~Edsjo, L.~Bergstrom, C.~Farnier and Y.~Akrami,
  JCAP {\bf 1001}, 031 (2010)
  [arXiv:0909.3300 [astro-ph.CO]].
  
\bibitem{Roszkowski:2012uf} 
  L.~Roszkowski, E.~M.~Sessolo and Y.~-L.~S.~Tsai,
  Phys.\ Rev.\ D {\bf 86}, 095005 (2012)
  [arXiv:1202.1503 [hep-ph]].





\bibitem{CMS:2012gu} 
  S.~Chatrchyan {\it et al.}  [CMS Collaboration],
  Phys.\ Lett.\ B {\bf 716}, 30 (2012)
  [arXiv:1207.7235 [hep-ex]].

\bibitem{ATLAS:2012gk} 
  G.~Aad {\it et al.}  [ATLAS Collaboration],
  Phys.\ Lett.\ B {\bf 716}, 1 (2012)
  [arXiv:1207.7214 [hep-ex]].    
  
  
\bibitem{multinest}
\url{http://projects.hepforge.org/multinest/}  
  
\bibitem{Kowalska:2012gs} 
  K.~Kowalska, S.~Munir, L.~Roszkowski, E.~M.~Sessolo, S.~Trojanowski and Y.~-L.~S.~Tsai,
  arXiv:1211.1693 [hep-ph].
  
\bibitem{Bechtle:2011sb} 
  P.~Bechtle, O.~Brein, S.~Heinemeyer, G.~Weiglein and K.~E.~Williams,
  Comput.\ Phys.\ Commun.\  {\bf 182}, 2605 (2011)
  [arXiv:1102.1898 [hep-ph]].
  
\bibitem{darksusy}
P.~Gondolo, J.~Edsjo, P.~Ullio, L.~Bergstrom, M.~Schelke and E.~A.~Baltz, JCAP 0407, 008 
(2004) [arXiv:astro-ph/0406204].

\bibitem{Cirelli:2007xd} 
  M.~Cirelli, A.~Strumia and M.~Tamburini,
  Nucl.\ Phys.\ B {\bf 787}, 152 (2007)
  [arXiv:0706.4071 [hep-ph]].

\bibitem{Hisano:2004ds}
  J.~Hisano, S.~Matsumoto, M.~M.~Nojiri and O.~Saito,
  Phys.\ Rev.\  D {\bf 71}, 063528 (2005).
  [arXiv:hep-ph/0412403].

\bibitem{Hryczuk:2010zi} 
  A.~Hryczuk, R.~Iengo and P.~Ullio,
  JHEP {\bf 1103}, 069 (2011)
  [arXiv:1010.2172 [hep-ph]].
 

\bibitem{Sjostrand:2006za}
  T.~Sjostrand, S.~Mrenna and P.~Skands,
  JHEP {\bf 0605}, 026 (2006).
  [arXiv:hep-ph/0603175].

\bibitem{Sjostrand:2007gs} 
  T.~Sjostrand, S.~Mrenna and P.~Z.~Skands,
  Comput.\ Phys.\ Commun.\  {\bf 178}, 852 (2008)
  [arXiv:0710.3820 [hep-ph]].

\bibitem{Corcella:2000bw} 
  G.~Corcella, I.~G.~Knowles, G.~Marchesini, S.~Moretti, K.~Odagiri, P.~Richardson, M.~H.~Seymour and B.~R.~Webber,
  JHEP {\bf 0101}, 010 (2001)
  [hep-ph/0011363].

\bibitem{Aprile:2012nq} 
  E.~Aprile {\it et al.}  [XENON100 Collaboration],
  Phys.\ Rev.\ Lett.\  {\bf 109}, 181301 (2012)
  [arXiv:1207.5988 [astro-ph.CO]].

\bibitem{likeCode}
\url{http://cosmology.bao.ac.cn/~huangxy/GetLdSphs_v1.tar.gz}


\end{thebibliography}
\end{document}